\documentclass[sigconf]{acmart} 

\usepackage{booktabs}
\usepackage{graphicx, epstopdf}
\usepackage{subfigure}
\usepackage{url}
\usepackage{multirow}
\usepackage{algorithm}
\usepackage{algorithmicx}
\usepackage{algcompatible}
\usepackage{balance}
\usepackage{arydshln}
\usepackage[normalem]{ulem}

\renewcommand*{\textrightarrow}{$\rightarrow$}

\copyrightyear{2025}
\acmYear{2025}
\setcopyright{acmlicensed}\acmConference[SIGIR '25]{Proceedings of the 48th International ACM SIGIR Conference on Research and Development in Information Retrieval}{July 13--18, 2025}{Padua, Italy}
\acmBooktitle{Proceedings of the 48th International ACM SIGIR Conference on Research and Development in Information Retrieval (SIGIR '25), July 13--18, 2025, Padua, Italy}
\acmDOI{10.1145/3726302.3730188}
\acmISBN{979-8-4007-1592-1/2025/07}

\begin{document}

\title{ELEC: Efficient Large Language Model-Empowered Click-Through Rate Prediction}

\author{Rui Dong}
\affiliation{%
  \institution{Alibaba Group}
  \city{Beijing}
  \country{China}
}
\email{kailu.dr@alibaba-inc.com}

\author{Wentao Ouyang}
\authornote{Corresponding author.}
\affiliation{%
  \institution{Alibaba Group}
  \city{Beijing}
   \country{China}
}
\email{maiwei.oywt@alibaba-inc.com}

\author{Xiangzheng Liu}
\affiliation{%
  \institution{Alibaba Group}
  \city{Beijing}
  \country{China}
}
\email{xiangzheng.lxz@alibaba-inc.com}

\begin{abstract}
Click-through rate (CTR) prediction plays an important role in online advertising systems. On the one hand, traditional CTR prediction models capture the collaborative signals in tabular data via feature interaction modeling, but they lose semantics in text. On the other hand, Large Language Models (LLMs) excel in understanding the context and meaning behind text, but they face challenges in capturing collaborative signals and they have long inference latency. In this paper, we aim to leverage the benefits of both types of models and pursue collaboration, semantics and efficiency. We present ELEC, which is an \underline{E}fficient \underline{L}LM-\underline{E}mpowered \underline{C}TR prediction framework. We first adapt an LLM for the CTR prediction task. In order to leverage the ability of the LLM but simultaneously keep efficiency, we utilize the pseudo-siamese network which contains a gain network and a vanilla network. We inject the high-level representation vector generated by the LLM into a collaborative CTR model to form the gain network such that it can take advantage of both tabular modeling and textual modeling. However, its reliance on the LLM limits its efficiency. We then distill the knowledge from the gain network to the vanilla network on both the score level and the representation level, such that the vanilla network takes only tabular data as input, but can still generate comparable performance as the gain network. Our approach is model-agnostic. It allows for the integration with various existing LLMs and collaborative CTR models. Experiments on real-world datasets demonstrate the effectiveness and efficiency of ELEC for CTR prediction.
\end{abstract}



\settopmatter{printacmref=true}

\maketitle

\section{Introduction}
Click-through rate (CTR) prediction is one of the most central tasks in online advertising systems \cite{he2014practical,shan2016deep,cheng2016wide,qu2016product,zhang2016deep,shah2017practical,wang2017deep,zhou2018deep,ouyang2019deep,qin2020user,tian2023eulernet}.
It aims to predict the probability that a user will click on a specific ad.
Traditional CTR prediction models heavily depend on tabular data (e.g., categorical features) and they capture the collaborative signals via feature interaction modeling, e.g., FM \cite{rendle2010factorization}, DeepFM \cite{guo2017deepfm}, AutoInt \cite{song2019autoint}, DCN \cite{wang2017deep} and DCNv2 \cite{wang2021dcn}. However, they cannot effectively process textual information.

Recently, the emergence of Large Language Models (LLMs) has given rise to another possibility for CTR prediction, which takes text as input. LLMs excel in understanding the context and meaning behind text, but they face challenges in capturing collaborative signals. In the recommendation domain \cite{zheng2024adapting}, language-based models show effectiveness in scenarios with limited user-item interactions, but they are inferior to traditional collaborative models when abundant user-item interactions are available \cite{yuan2023go,bao2023tallrec,kim2024large}. Moreover, LLMs suffer from low efficiency in online inference.

In this paper, we aim to leverage the benefits of both types of models and pursue collaboration, semantics and efficiency. We present ELEC, which is an \underline{E}fficient \underline{L}LM-\underline{E}mpowered \underline{C}TR prediction framework. In order to tailor an LLM for CTR prediction, we add a transformation layer and a prediction layer, and train the newly added parameters on CTR data.
In order to leverage the ability of the LLM but maintain efficiency at the same time, we utilize the pseudo-siamese network which contains a gain network and a vanilla network. We inject the high-level representation vector generated by the LLM into a collaborative model to form the gain network such that it can benefit from both tabular modeling and textual modeling. However, the gain network is not efficient. We then distill the knowledge from the gain network to the vanilla network such that the latter takes only tabular data as input, but can still generate comparable performance as the former.

The main contributions of this paper are
\begin{itemize}
\item We propose ELEC for CTR prediction. ELEC leverages the benefits of both collaborative models and language models. It thus captures both collaborative and semantic signals.
\item ELEC is efficient. Its online part does not depend on LLMs and eliminates the long inference latency by LLMs.
\item ELEC is also model-agnostic. It allows the integration of various existing LLMs and collaborative CTR prediction models.
\item We conduct experiments on real-world datasets to evaluate the performance of various types of methods.
\end{itemize}

\begin{figure*}[!t]
\vskip -10pt
\centering
\includegraphics[width=\textwidth, trim = 0 0 0 0, clip]{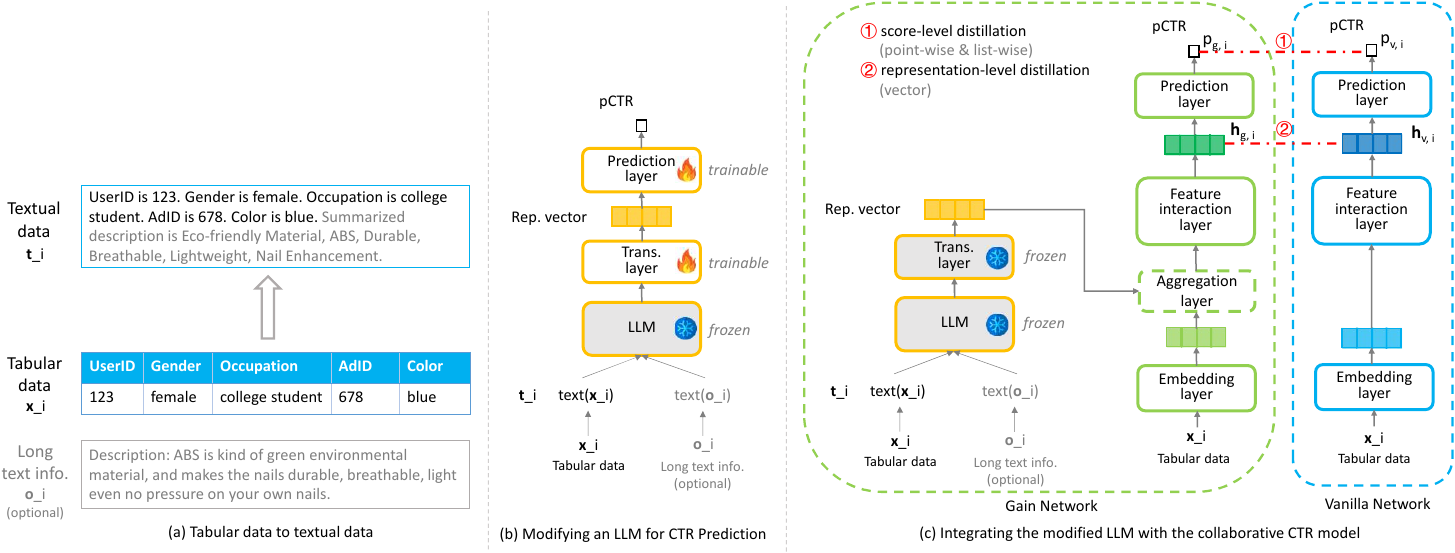}
\vskip -10pt
\caption{ELEC framework. (a) Tabular data to textual data. (b) Modifying an LLM for CTR prediction (Trans. - Transformation; Rep. - Representation). (c) Integrating the modified LLM with the collaborative CTR model. For simplicity, the aggregation layer in the gain network uses vector concatenation. Only the \textnormal{vanilla network} is used for \textnormal{online inference}.}
\vskip -10pt
\label{fig_model}
\end{figure*}

\section{Model Design}

\subsection{Background}
We denote the CTR dataset as ${(\mathbf{x}_i , \mathbf{t}_i, y_i)}_{i=1}^M$, where $i$ is the sample index, $\mathbf{x}_i$ is tabular data, $\mathbf{t}_i$ is textual data and $y_i \in \{0,1\}$ is the label indicating a user's actual click behavior. The goal of CTR prediction is to estimate the click probability based on the given input.

Traditional CTR prediction models take tabular data $\mathbf{x}_i$ as input and generate a probability in $[0,1]$ as output.
LLMs are not designed for the CTR prediction task, they can take textual data $\mathbf{t}_i$ as input but they generate text as output.

Our goal is to leverage the benefits of both types of models and simultaneously keep the model efficient for CTR prediction.

\subsection{Modifying an LLM for CTR Prediction} \label{sec}
LLMs provide a potential way for CTR prediction by leveraging their powerful knowledge and textual reasoning ability.
However, LLMs are not initially designed for CTR prediction.
We make the following three revisions in order to adapt an LLM for CTR prediction.

\subsubsection{\textbf{Input: from tabular data to textual data}}
The input for CTR prediction consists of tabular data (e.g., categorical features) that cannot be directly processed by LLMs.
Therefore, we first transform tabular data into textual data. We adopt a simple yet effective transformation template \cite{li2025ctrl}, which is the concatenation of ``<\texttt{field\_name}> is <\texttt{feature\_value}>''  (Fig. \ref{fig_model}(a)).
For example, if the original tabular data $\mathbf{x}_i$ is ``gender: female, occupation: college student'', then the transformed text is ``Gender is female. Occupation is college student.''
Moreover, other text information $\mathbf{o}_i$ such as long descriptions (optional) that are not in $\mathbf{x}_i$ can also be fed to the LLM. To reduce noise in long text, we can use any LLM to summarize the long text into tags. All such text information forms the input $\mathbf{t}_i$ to the LLM.

\subsubsection{\textbf{Output: adding a transformation layer and a prediction layer}}
Given the set of hidden states from the last layer of the LLM, we take the \textbf{average pooling} of these hidden states as the initial semantic information representation.
We then add a transformation layer to transform this representation from the semantic space to the CTR prediction space (Fig. \ref{fig_model}(b)). For simplicity, the transformation layer consists of an MLP with several fully connected layers. We then obtain a high-level representation vector.
Finally, a prediction layer with the sigmoid activation function is used to output the CTR prediction (i.e., a value in $[0,1]$).

\subsubsection{\textbf{Training: training newly added model parameters}}
We keep the original model parameters of the LLM frozen and only train the newly added model parameters in the transformation layer and the prediction layer using CTR data. In this way, the modified LLM is adapted to the CTR prediction task.

\subsection{Integrating the Modified LLM and the Traditional Collaborative CTR Model}
\subsubsection{\textbf{Design Objectives}}
We would like the target model to be both effective and efficient. By effective, we mean that the target model should integrate the strength from both the modified LLM and the traditional collaborative CTR model. By efficient, we mean that the target model should have short online inference time.

\subsubsection{\textbf{Pseudo-siamese Network}}
We leverage the pseudo-siamese network to achieve both effectiveness and efficiency. The pseudo-siamese network offers more flexibility than the siamese network, as it allows two branches to have different structures and to receive inputs from various modalities \cite{wang2024adsnet}.
In particular, our pseudo-siamese network consists of a gain network and a vanilla network.
The gain network has a modified LLM module (all parameters are frozen) and a collaborative module, while the vanilla network has only a collaborative module (Fig. \ref{fig_model}(c)).
The collaborative modules of the gain network and the vanilla network have the same structure, e.g., both are DNN \cite{cheng2016wide} or DCNv2 \cite{wang2021dcn}.
Generally, collaborative CTR models adopt the ``Embedding \& Feature Interaction'' paradigm \cite{song2019autoint,wang2021dcn}.
The input to the gain network contains both textual data $\mathbf{t}_i$ and tabular data $\mathbf{x}_i$, while the input to the vanilla network contains only tabular data $\mathbf{x}_i$.

\subsubsection{\textbf{Model training and knowledge distillation}}
The gain network is expected to generate better performance as it can process both textual data and tabular data. However, it is not efficient.
The vanilla network is efficient by design, but it can process only tabular data.
Therefore, we would like to transfer the knowledge from the gain network to the vanilla network.
In this way, the vanilla network achieves both effectiveness and efficiency.

We perform knowledge distillation at two levels: \textbf{score level} (coarse-grained scalar) and \textbf{representation level} (fine-grained vector). At the score level, we would like the vanilla network not only mimics the point-wise calibration ability of the gain network, but also the list-wise ranking ability. Therefore, we utilize the Calibration-compatible LIstwise Distillation (CLID) loss \cite{gui2024calibration} to achieve this goal. It is given by
\begin{equation}
L_{score} = - \frac{1}{N} \sum_{i=1}^N Q_{g,i} \log Q_{v,i},
\end{equation}
\begin{equation}
Q_{g,i} = \frac{p_{g,i}}{\sum_{j=1}^N p_{g,j}}, \ \ Q_{v,i} = \frac{p_{v,i}}{\sum_{j=1}^N p_{v,j}},
\end{equation}
where $i$ is the sample index, $N$ is the number of samples in a mini-batch, $p_{g,i}$ is the estimated probability by the gain network and $p_{v,i}$ is the estimated probability by the vanilla network.

At the representation level, we would like the high-level representation vector $\mathbf{h}_{v, i}$ of the vanilla network to be similar to $\mathbf{h}_{g, i}$ of the gain network (Fig. \ref{fig_model}(c)).
In particular, we use the mean squared error (MSE) loss to achieve this goal
\begin{equation}
L_{rep} = \frac{1}{N} \sum_{i=1}^N \left \|\mathbf{h}_{v, i} - \mathbf{h}_{g, i} \right\|^2.
\end{equation}

Moreover, both networks use the label information $y_i$ to guide the training and optimize the binary cross-entropy loss
\begin{align}
&L_{gain} = - \frac{1}{N} \sum_{i=1}^N \left(y_i \log p_{g,i} + (1 - y_i) \log(1 - p_{g,i}) \right) \nonumber \\
&L_{van} = - \frac{1}{N} \sum_{i=1}^N \left(y_i \log p_{v,i} + (1 - y_i) \log(1 - p_{v,i}) \right).
\end{align}

Finally, we train the gain network and the vanilla network jointly
\begin{equation}
L_{total} = L_{gain} + L_{van} + L_{score}  + \alpha L_{rep},
\end{equation}
where $\alpha$ is a balancing hyperparameter. Only collaborative modules are trained. $L_{score}$ and $L_{rep}$ only impact the vanilla network.

\subsubsection{\textbf{Online Inference}}
We use the vanilla network for online inference, which takes only tabular data as input.
As the performance of the vanilla network is similar to that of the gain network, but the vanilla network does not depend on the LLM and has only a collaborative module, it achieves both effectiveness and efficiency.

\section{Experiments}

\subsection{Datasets}
The statistics of the experimental datasets are listed in Table \ref{tab_stat}.

1) \textbf{Amazon dataset} \cite{hou2024bridging}. We use a subset ``All Beauty''. We convert ratings that are greater than 3 to label 1 and others to label 0.

2) \textbf{Industry dataset}. It is sampled from the click log of an industrial news feed advertising platform.

\begin{table}[!t]
\setlength{\tabcolsep}{2pt}
\renewcommand{\arraystretch}{1.1}
\caption{Statistics of experimental datasets.}
\vskip -11pt
\label{tab_stat}
\centering
\begin{tabular}{|l|c|c|c|c|c|c|}
\hline
\textbf{Dataset} & \textbf{\# Fields} & \textbf{\# Train} & \textbf{\# Val} & \textbf{\# Test} & \textbf{\# Show} & \textbf{\# Click} \\
\hline
Amazon & 8 & 67.80M & 8.47M & 8.48M & 84.75M & 60.50M \\
\hline
Industry & 32 & 571.0M & 81.3M & 80.2M & 732.5M & 284.7M \\
\hline
\end{tabular}
\vskip -12pt
\end{table}

\subsection{Methods in Comparison}
We compare the following three categories of models for CTR prediction.
1) \emph{Collaborative models}.
\begin{itemize}
\item \textbf{DNN} \cite{cheng2016wide}. Deep Neural Network. It contains an embedding layer, several fully connected layers and an output layer.
\item \textbf{DeepFM} \cite{guo2017deepfm}. It combines factorization machine and DNN.
\item \textbf{AutoInt} \cite{song2019autoint}. It consists of a multi-head self-attentive network with residual connections and DNN.
\item \textbf{DCNv2} \cite{wang2021dcn}. Improved Deep \& Cross Network.
\end{itemize}

2) \emph{Language models}.
\begin{itemize}
\item \textbf{CTR-BERT} \cite{muhamed2021ctr}. It adopts two-tower BERT \cite{devlin2018bert} and feeds the semantic information of user and item separately to get the prediction score.
\item \textbf{P5} \cite{geng2022recommendation}. It converts various recommendation tasks into text generation tasks by prompt learning.
\item \textbf{M-LLM}. The modified LLM in \S\ref{sec}.
\end{itemize}

3) \emph{Language model-empowered collaborative models}.
\begin{itemize}
\item \textbf{CTRL} \cite{li2025ctrl}. It considers tabular data and transformed textual data as two different modalities, uses contrastive learning between them to pre-train a collaborative model, and then fine-tunes the model for CTR prediction.
\item \textbf{FLIP} \cite{wang2024flip}. It learns fine-grained alignment between tabular data and transformed textual data to pre-train a collaborative model, and then fine-tunes the model for CTR prediction.
\item \textbf{ELEC}. The proposed efficient LLM-empowered CTR prediction framework in this paper.
\end{itemize}

All the above methods are based on standard CTR features per sample.
For fair comparison, we do not include methods such as DIN \cite{zhou2018deep} and DIEN \cite{zhou2019deep} that take additional information (e.g., historical user behaviors) into consideration.

\subsection{Settings}
\textbf{Parameters.} We set the embedding dimension as 32. CTRL, FLIP and ELEC use gte-Qwen2-7B-instruct as the LLM. ELEC uses DCNv2 \cite{wang2021dcn} as the collaborative model. The number of cross layers is 2 and layer dimensions in the deep part are \{256, 128, 64\}. Dimensions in the transformation layer of M-LLM are \{512, 256, 128\}.

\textbf{Evaluation Metric.} We use AUC and LogLoss as the evaluation metrics for CTR prediction \cite{song2019autoint,wang2021dcn}.

\subsection{Experimental Results}
\subsubsection{\textbf{Effectiveness}}
\begin{table}[!t]
\setlength{\tabcolsep}{2pt}
\renewcommand{\arraystretch}{1.1}
\caption{Test results on experimental datasets. The best result is in bold font. A \textnormal{small} improvement in AUC (e.g., \textnormal{0.0020}) can lead to a \textnormal{significant} increase in online CTR (e.g., 3\%) \cite{cheng2016wide}. * indicates the statistical significance for $p \leq 0.01$ compared with the second best result over paired t-test.}
\vskip -10pt
\label{tab_auc}
\centering
\begin{tabular}{|l|c c|c c|}
\hline
 & \multicolumn{2}{|c|}{\textbf{Amazon}} & \multicolumn{2}{|c|}{\textbf{Industry}} \\
\hline
& AUC$\uparrow$ & LogLoss$\downarrow$ & AUC$\uparrow$ & LogLoss$\downarrow$  \\
\hline
DNN                 & 0.7408 & 0.5201 & 0.7922 & 0.4843 \\
DeepFM          & 0.7413 & 0.5181 & 0.7926 & 0.4844 \\
AutoInt            & 0.7421 & 0.5172 & 0.7935 & 0.4825 \\
DCNv2              & 0.7430 & 0.5166 & 0.7946 & 0.4812 \\
\hline
CTR-BERT & 0.7095 & 0.5373 & 0.7602 & 0.4980 \\
P5               & 0.7102 & 0.5359 & 0.7591 & 0.4991 \\
M-LLM      & 0.7115 & 0.5347 & 0.7617 & 0.4963 \\
\hline
CTRL & 0.7475 & 0.5103 & 0.7973 & 0.4768 \\
FLIP  & 0.7489 & 0.5087 & 0.7981 & 0.4756 \\
ELEC & \textbf{0.7509}* & \textbf{0.5051}* & \textbf{0.8015}* & \textbf{0.4694}* \\
\hline
\end{tabular}
\vskip -8pt
\end{table}

It is observed in Table \ref{tab_auc} that language models generally perform worse than collaborative models for CTR prediction, which complies with previous studies \cite{li2025ctrl}. It is probably because language models do not explicitly capture the collaborative signals in CTR features. However, language models have their own understanding about CTR data and provide complementary information. As a result, language model-empowered collaborative models such as CTRL, FLIP and ELEC achieve better performance than other methods.
CTRL and FLIP use the original language model without adaptation. They pre-train a collaborative model and then fine-tune it.
ELEC adds a transformation layer and a prediction layer after the original language model in order to adapt it for better CTR prediction. ELEC explicitly injects the high-level representation vector of the language model into a collaborative model. ELEC achieves the highest AUCs on experimental datasets.

\subsubsection{\textbf{Ablation Study}}

\begin{table}[!t]
\setlength{\tabcolsep}{2pt}
\renewcommand{\arraystretch}{1.1}
\caption{Ablation study.}
\vskip -10pt
\label{tab_ab}
\centering
\begin{tabular}{|l|c c|c c|}
\hline
 & \multicolumn{2}{|c|}{\textbf{Amazon}} & \multicolumn{2}{|c|}{\textbf{Industry}} \\
\hline
& AUC$\uparrow$ & LogLoss$\downarrow$ & AUC$\uparrow$ & LogLoss$\downarrow$  \\
\hline
wo textual input & 0.7430 & 0.5166 & 0.7946 & 0.4812 \\
wo CLID loss & 0.7487 & 0.5095 & 0.7992 & 0.4760 \\
wo MSE loss & 0.7496 & 0.5072 & 0.8006 & 0.4708 \\
ELEC & \textbf{0.7509} & \textbf{0.5051} & \textbf{0.8015} & \textbf{0.4694} \\
\hline
\end{tabular}
\vskip -8pt
\end{table}

It is observed in Table \ref{tab_ab} that without textual input, without the CLID loss and without the MSE loss all lead to worse performance. It shows the effectiveness of these model components. Without the CLID loss leads to large performance degradation, showing the effectiveness of the CLID loss, which is capable of distilling both the calibration ability and the ranking ability of the gain network to the vanilla network.

\subsubsection{\textbf{Effect of Base LLMs}}
\begin{table}[!t]
\setlength{\tabcolsep}{2pt}
\renewcommand{\arraystretch}{1.1}
\caption{Effect of base LLMs.}
\vskip -10pt
\label{tab_llm}
\centering
\begin{tabular}{|l|c c|c c|}
\hline
 & \multicolumn{2}{|c|}{\textbf{Amazon}} & \multicolumn{2}{|c|}{\textbf{Industry}} \\
\hline
Base LLM in ELEC & AUC$\uparrow$ & LogLoss$\downarrow$ & AUC$\uparrow$ & LogLoss$\downarrow$  \\
\hline
roberta-base & 0.7481 & 0.5084  & 0.7995 & 0.4710 \\
chatglm-6B & 0.7492 & 0.5069 & 0.8004 & 0.4704 \\
gte-Qwen2-7B-instruct & \textbf{0.7509} & \textbf{0.5051} & \textbf{0.8015} & \textbf{0.4694} \\
\hline
\end{tabular}
\vskip -8pt
\end{table}

Table \ref{tab_llm} shows the performance of ELEC with different language models.
Roberta has the fewest number of model parameters and Qwen2-7B has the largest number of model parameters.
ELEC with Roberta achieves the lowest AUC while ELEC with Qwen2-7B achieves the highest AUC.
These results show that more sophisticated language models generally lead to better performance.

\subsubsection{\textbf{Efficiency}}

\begin{table}[!t]
\setlength{\tabcolsep}{1pt}
\renewcommand{\arraystretch}{1}
\caption{Inference time per sample.}
\vskip -10pt
\label{tab_time}
\centering
\begin{tabular}{|l|c c c|c c c|}
\hline
 & \multicolumn{3}{|c|}{\textbf{Amazon}} & \multicolumn{3}{|c|}{\textbf{Industry}} \\
\hline
{\small Base LLM in ELEC} & {\small M-LLM} &  {\small gain net.} & {\small van. net.} & {\small M-LLM} & {\small gain net.} & {\small van. net.} \\
\hline
{\small roberta-base} & 0.164s & 0.186s & \textbf{0.087s} & 0.194s & 0.238s & \textbf{0.116s} \\
{\small chatglm-6B} &1.113s & 1.138s & \textbf{0.087s} & 1.311s & 1.359s & \textbf{0.116s} \\
{\small gte-Qwen2-7B-instruct} & 0.715s & 0.734s & \textbf{0.087s} & 0.977s & 1.023s & \textbf{0.116s} \\
\hline
\end{tabular}
\vskip -8pt
\end{table}

Table \ref{tab_time} lists the online inference time per sample using NVIDIA V100.
Qwen2-7B is optimized for inference, and its inference time is shorter than ChatGLM-6B.
It is observed that the gain network has the longest inference time because it integrates a modified LLM and a collaborative model. The vanilla network has the smallest inference time as it consists of only a collaborative model. This time is independent of the base LLM used, which means that the performance of ELEC can be improved by using more advanced LLMs, but the inference remains fast.

\section{Related Work}
\textbf{Collaborative models for CTR prediction.}
CTR prediction has attracted lots of attention in recent years. Methods range from shallow models such as LR \cite{richardson2007predicting}, FM \cite{rendle2010factorization},  Field-weighted FM \cite{pan2018field} to deep models such as Wide\&Deep \cite{cheng2016wide}, DeepFM \cite{guo2017deepfm}, xDeepFM \cite{lian2018xdeepfm}, AutoInt \cite{song2019autoint}, DCNv2 \cite{wang2021dcn} and DIL \cite{zhang2023reformulating}. Deep models generally adopt the ``Embedding \& Feature Interaction'' paradigm.

\textbf{Language models for CTR prediction.}
CTR-BERT \cite{muhamed2021ctr} has a two-tower structure with two BERT models, which encode user and item text information respectively.
P5 \cite{geng2022recommendation} considers recommendation as language processing. All data such as user-item interactions, user descriptions, item metadata, and user reviews are converted to natural language sequences.
These models could be modified for CTR prediction. However, they cannot effectively model collaborative signals and generate suboptimal performance. Moreover, they suffer from low efficiency for online inference.

\textbf{Language model-empowered collaborative models for CTR prediction.}
CTRL \cite{li2025ctrl}, FLIP \cite{wang2024flip} and ELEC fall in this category. CTRL performs contrastive learning between tabular data and textual data to pre-train a collaborative model, and then fine-tunes it for CTR prediction. FLIP performs more fine-grained pre-training. CTRL and FLIP use the original language model. Differently, ELEC first adapts an LLM for CTR prediction by adding a transformation layer and a prediction layer, and then training the newly-added parameters on CTR data. It then injects the high-level representation vector generated by the LLM into a collaborative CTR model to form a gain network, and distills the knowledge to a lightweight vanilla network to achieve both effectiveness and efficiency.

\section{Conclusion}
In this paper, we propose ELEC for CTR prediction. ELEC leverages the benefits of both collaborative models and language models. It is able to capture both collaborative and semantic signals. ELEC is model-agnostic. It allows the integration of various existing LLMs and collaborative CTR models. ELEC is also efficient. The online inference time is independent of the base LLM used, which means the performance of ELEC can be improved by using more advanced LLMs, but the inference remains fast. Experimental results demonstrate the effectiveness and efficiency of ELEC for CTR prediction.

\bibliographystyle{ACM-Reference-Format}
\balance
\bibliography{ref}


\begin{thebibliography}{32}


\ifx \showCODEN    \undefined \def \showCODEN     #1{\unskip}     \fi
\ifx \showISBNx    \undefined \def \showISBNx     #1{\unskip}     \fi
\ifx \showISBNxiii \undefined \def \showISBNxiii  #1{\unskip}     \fi
\ifx \showISSN     \undefined \def \showISSN      #1{\unskip}     \fi
\ifx \showLCCN     \undefined \def \showLCCN      #1{\unskip}     \fi
\ifx \shownote     \undefined \def \shownote      #1{#1}          \fi
\ifx \showarticletitle \undefined \def \showarticletitle #1{#1}   \fi
\ifx \showURL      \undefined \def \showURL       {\relax}        \fi
\providecommand\bibfield[2]{#2}
\providecommand\bibinfo[2]{#2}
\providecommand\natexlab[1]{#1}
\providecommand\showeprint[2][]{arXiv:#2}

\bibitem[Bao et~al\mbox{.}(2023)]%
        {bao2023tallrec}
\bibfield{author}{\bibinfo{person}{Keqin Bao}, \bibinfo{person}{Jizhi Zhang},
  \bibinfo{person}{Yang Zhang}, \bibinfo{person}{Wenjie Wang},
  \bibinfo{person}{Fuli Feng}, {and} \bibinfo{person}{Xiangnan He}.}
  \bibinfo{year}{2023}\natexlab{}.
\newblock \showarticletitle{Tallrec: An effective and efficient tuning
  framework to align large language model with recommendation}. In
  \bibinfo{booktitle}{\emph{Proceedings of the 17th ACM Conference on
  Recommender Systems (RecSys)}}. \bibinfo{pages}{1007--1014}.
\newblock


\bibitem[Cheng et~al\mbox{.}(2016)]%
        {cheng2016wide}
\bibfield{author}{\bibinfo{person}{Heng-Tze Cheng}, \bibinfo{person}{Levent
  Koc}, \bibinfo{person}{Jeremiah Harmsen}, \bibinfo{person}{Tal Shaked},
  \bibinfo{person}{Tushar Chandra}, \bibinfo{person}{Hrishi Aradhye},
  \bibinfo{person}{Glen Anderson}, \bibinfo{person}{Greg Corrado},
  \bibinfo{person}{Wei Chai}, \bibinfo{person}{Mustafa Ispir}, {et~al\mbox{.}}}
  \bibinfo{year}{2016}\natexlab{}.
\newblock \showarticletitle{Wide \& deep learning for recommender systems}. In
  \bibinfo{booktitle}{\emph{Proceedings of the 1st Workshop on Deep Learning
  for Recommender Systems (DLRS)}}. ACM, \bibinfo{pages}{7--10}.
\newblock


\bibitem[Devlin(2018)]%
        {devlin2018bert}
\bibfield{author}{\bibinfo{person}{Jacob Devlin}.}
  \bibinfo{year}{2018}\natexlab{}.
\newblock \showarticletitle{Bert: Pre-training of deep bidirectional
  transformers for language understanding}.
\newblock \bibinfo{journal}{\emph{arXiv preprint arXiv:1810.04805}}
  (\bibinfo{year}{2018}).
\newblock


\bibitem[Geng et~al\mbox{.}(2022)]%
        {geng2022recommendation}
\bibfield{author}{\bibinfo{person}{Shijie Geng}, \bibinfo{person}{Shuchang
  Liu}, \bibinfo{person}{Zuohui Fu}, \bibinfo{person}{Yingqiang Ge}, {and}
  \bibinfo{person}{Yongfeng Zhang}.} \bibinfo{year}{2022}\natexlab{}.
\newblock \showarticletitle{Recommendation as language processing (RLP): A
  unified pretrain, personalized prompt \& predict paradigm (P5)}. In
  \bibinfo{booktitle}{\emph{Proceedings of the 16th ACM Conference on
  Recommender Systems (RecSys)}}. \bibinfo{pages}{299--315}.
\newblock


\bibitem[Gui et~al\mbox{.}(2024)]%
        {gui2024calibration}
\bibfield{author}{\bibinfo{person}{Xiaoqiang Gui}, \bibinfo{person}{Yueyao
  Cheng}, \bibinfo{person}{Xiang-Rong Sheng}, \bibinfo{person}{Yunfeng Zhao},
  \bibinfo{person}{Guoxian Yu}, \bibinfo{person}{Shuguang Han},
  \bibinfo{person}{Yuning Jiang}, \bibinfo{person}{Jian Xu}, {and}
  \bibinfo{person}{Bo Zheng}.} \bibinfo{year}{2024}\natexlab{}.
\newblock \showarticletitle{Calibration-compatible Listwise Distillation of
  Privileged Features for CTR Prediction}. In
  \bibinfo{booktitle}{\emph{Proceedings of the 17th ACM International
  Conference on Web Search and Data Mining (WSDM)}}. \bibinfo{pages}{247--256}.
\newblock


\bibitem[Guo et~al\mbox{.}(2017)]%
        {guo2017deepfm}
\bibfield{author}{\bibinfo{person}{Huifeng Guo}, \bibinfo{person}{Ruiming
  Tang}, \bibinfo{person}{Yunming Ye}, \bibinfo{person}{Zhenguo Li}, {and}
  \bibinfo{person}{Xiuqiang He}.} \bibinfo{year}{2017}\natexlab{}.
\newblock \showarticletitle{Deepfm: a factorization-machine based neural
  network for ctr prediction}. In \bibinfo{booktitle}{\emph{International Joint
  Conferences on Artificial Intelligence (IJCAI)}}.
  \bibinfo{pages}{1725--1731}.
\newblock


\bibitem[He et~al\mbox{.}(2014)]%
        {he2014practical}
\bibfield{author}{\bibinfo{person}{Xinran He}, \bibinfo{person}{Junfeng Pan},
  \bibinfo{person}{Ou Jin}, \bibinfo{person}{Tianbing Xu}, \bibinfo{person}{Bo
  Liu}, \bibinfo{person}{Tao Xu}, \bibinfo{person}{Yanxin Shi},
  \bibinfo{person}{Antoine Atallah}, \bibinfo{person}{Ralf Herbrich},
  \bibinfo{person}{Stuart Bowers}, {et~al\mbox{.}}}
  \bibinfo{year}{2014}\natexlab{}.
\newblock \showarticletitle{Practical lessons from predicting clicks on ads at
  facebook}. In \bibinfo{booktitle}{\emph{Proceedings of the International
  Workshop on Data Mining for Online Advertising (ADKDD)}}. ACM,
  \bibinfo{pages}{1--9}.
\newblock


\bibitem[Hou et~al\mbox{.}(2024)]%
        {hou2024bridging}
\bibfield{author}{\bibinfo{person}{Yupeng Hou}, \bibinfo{person}{Jiacheng Li},
  \bibinfo{person}{Zhankui He}, \bibinfo{person}{An Yan},
  \bibinfo{person}{Xiusi Chen}, {and} \bibinfo{person}{Julian McAuley}.}
  \bibinfo{year}{2024}\natexlab{}.
\newblock \showarticletitle{Bridging Language and Items for Retrieval and
  Recommendation}.
\newblock \bibinfo{journal}{\emph{arXiv preprint arXiv:2403.03952}}
  (\bibinfo{year}{2024}).
\newblock


\bibitem[Kim et~al\mbox{.}(2024)]%
        {kim2024large}
\bibfield{author}{\bibinfo{person}{Sein Kim}, \bibinfo{person}{Hongseok Kang},
  \bibinfo{person}{Seungyoon Choi}, \bibinfo{person}{Donghyun Kim},
  \bibinfo{person}{Minchul Yang}, {and} \bibinfo{person}{Chanyoung Park}.}
  \bibinfo{year}{2024}\natexlab{}.
\newblock \showarticletitle{Large language models meet collaborative filtering:
  An efficient all-round llm-based recommender system}. In
  \bibinfo{booktitle}{\emph{Proceedings of the 30th ACM SIGKDD Conference on
  Knowledge Discovery and Data Mining (KDD)}}. \bibinfo{pages}{1395--1406}.
\newblock


\bibitem[Li et~al\mbox{.}(2025)]%
        {li2025ctrl}
\bibfield{author}{\bibinfo{person}{Xiangyang Li}, \bibinfo{person}{Bo Chen},
  \bibinfo{person}{Lu Hou}, {and} \bibinfo{person}{Ruiming Tang}.}
  \bibinfo{year}{2025}\natexlab{}.
\newblock \showarticletitle{Ctrl: Connect collaborative and language model for
  ctr prediction}.
\newblock \bibinfo{journal}{\emph{ACM Transactions on Recommender Systems}}
  (\bibinfo{year}{2025}).
\newblock


\bibitem[Lian et~al\mbox{.}(2018)]%
        {lian2018xdeepfm}
\bibfield{author}{\bibinfo{person}{Jianxun Lian}, \bibinfo{person}{Xiaohuan
  Zhou}, \bibinfo{person}{Fuzheng Zhang}, \bibinfo{person}{Zhongxia Chen},
  \bibinfo{person}{Xing Xie}, {and} \bibinfo{person}{Guangzhong Sun}.}
  \bibinfo{year}{2018}\natexlab{}.
\newblock \showarticletitle{xDeepFM: Combining explicit and implicit feature
  interactions for recommender systems}. In
  \bibinfo{booktitle}{\emph{Proceedings of the 24th ACM SIGKDD International
  Conference on Knowledge Discovery \& Data Mining (KDD)}}. ACM,
  \bibinfo{pages}{1754--1763}.
\newblock


\bibitem[Muhamed et~al\mbox{.}(2021)]%
        {muhamed2021ctr}
\bibfield{author}{\bibinfo{person}{Aashiq Muhamed}, \bibinfo{person}{Iman
  Keivanloo}, \bibinfo{person}{Sujan Perera}, \bibinfo{person}{James Mracek},
  \bibinfo{person}{Yi Xu}, \bibinfo{person}{Qingjun Cui},
  \bibinfo{person}{Santosh Rajagopalan}, \bibinfo{person}{Belinda Zeng}, {and}
  \bibinfo{person}{Trishul Chilimbi}.} \bibinfo{year}{2021}\natexlab{}.
\newblock \showarticletitle{CTR-BERT: Cost-effective knowledge distillation for
  billion-parameter teacher models}. In \bibinfo{booktitle}{\emph{NeurIPS
  Efficient Natural Language and Speech Processing Workshop}}.
\newblock


\bibitem[Ouyang et~al\mbox{.}(2019)]%
        {ouyang2019deep}
\bibfield{author}{\bibinfo{person}{Wentao Ouyang}, \bibinfo{person}{Xiuwu
  Zhang}, \bibinfo{person}{Li Li}, \bibinfo{person}{Heng Zou},
  \bibinfo{person}{Xin Xing}, \bibinfo{person}{Zhaojie Liu}, {and}
  \bibinfo{person}{Yanlong Du}.} \bibinfo{year}{2019}\natexlab{}.
\newblock \showarticletitle{Deep spatio-temporal neural networks for
  click-through rate prediction}. In \bibinfo{booktitle}{\emph{Proceedings of
  the 25th ACM SIGKDD International Conference on Knowledge Discovery \& Data
  Mining (KDD)}}. ACM, \bibinfo{pages}{2078--2086}.
\newblock


\bibitem[Pan et~al\mbox{.}(2018)]%
        {pan2018field}
\bibfield{author}{\bibinfo{person}{Junwei Pan}, \bibinfo{person}{Jian Xu},
  \bibinfo{person}{Alfonso~Lobos Ruiz}, \bibinfo{person}{Wenliang Zhao},
  \bibinfo{person}{Shengjun Pan}, \bibinfo{person}{Yu Sun}, {and}
  \bibinfo{person}{Quan Lu}.} \bibinfo{year}{2018}\natexlab{}.
\newblock \showarticletitle{Field-weighted Factorization Machines for
  Click-Through Rate Prediction in Display Advertising}. In
  \bibinfo{booktitle}{\emph{Proceedings of the 2018 World Wide Web Conference
  (WWW)}}. IW3C2, \bibinfo{pages}{1349--1357}.
\newblock


\bibitem[Qin et~al\mbox{.}(2020)]%
        {qin2020user}
\bibfield{author}{\bibinfo{person}{Jiarui Qin}, \bibinfo{person}{Weinan Zhang},
  \bibinfo{person}{Xin Wu}, \bibinfo{person}{Jiarui Jin},
  \bibinfo{person}{Yuchen Fang}, {and} \bibinfo{person}{Yong Yu}.}
  \bibinfo{year}{2020}\natexlab{}.
\newblock \showarticletitle{User Behavior Retrieval for Click-Through Rate
  Prediction}. In \bibinfo{booktitle}{\emph{Proceedings of the 43rd
  International ACM SIGIR Conference on Research and Development in Information
  Retrieval (SIGIR)}}. ACM, \bibinfo{pages}{2347--2356}.
\newblock


\bibitem[Qu et~al\mbox{.}(2016)]%
        {qu2016product}
\bibfield{author}{\bibinfo{person}{Yanru Qu}, \bibinfo{person}{Han Cai},
  \bibinfo{person}{Kan Ren}, \bibinfo{person}{Weinan Zhang},
  \bibinfo{person}{Yong Yu}, \bibinfo{person}{Ying Wen}, {and}
  \bibinfo{person}{Jun Wang}.} \bibinfo{year}{2016}\natexlab{}.
\newblock \showarticletitle{Product-based neural networks for user response
  prediction}. In \bibinfo{booktitle}{\emph{2016 IEEE 16th International
  Conference on Data Mining (ICDM)}}. IEEE, \bibinfo{pages}{1149--1154}.
\newblock


\bibitem[Rendle(2010)]%
        {rendle2010factorization}
\bibfield{author}{\bibinfo{person}{Steffen Rendle}.}
  \bibinfo{year}{2010}\natexlab{}.
\newblock \showarticletitle{Factorization machines}. In
  \bibinfo{booktitle}{\emph{IEEE 10th International Conference on Data Mining
  (ICDM)}}. IEEE, \bibinfo{pages}{995--1000}.
\newblock


\bibitem[Richardson et~al\mbox{.}(2007)]%
        {richardson2007predicting}
\bibfield{author}{\bibinfo{person}{Matthew Richardson}, \bibinfo{person}{Ewa
  Dominowska}, {and} \bibinfo{person}{Robert Ragno}.}
  \bibinfo{year}{2007}\natexlab{}.
\newblock \showarticletitle{Predicting clicks: estimating the click-through
  rate for new ads}. In \bibinfo{booktitle}{\emph{Proceedings of the 16th
  International Conference on World Wide Web (WWW)}}. IW3C2,
  \bibinfo{pages}{521--530}.
\newblock


\bibitem[Shah et~al\mbox{.}(2017)]%
        {shah2017practical}
\bibfield{author}{\bibinfo{person}{Parikshit Shah}, \bibinfo{person}{Ming
  Yang}, \bibinfo{person}{Sachidanand Alle}, \bibinfo{person}{Adwait
  Ratnaparkhi}, \bibinfo{person}{Ben Shahshahani}, {and} \bibinfo{person}{Rohit
  Chandra}.} \bibinfo{year}{2017}\natexlab{}.
\newblock \showarticletitle{A practical exploration system for search
  advertising}. In \bibinfo{booktitle}{\emph{Proceedings of the 23rd ACM SIGKDD
  International Conference on Knowledge Discovery and Data Mining (KDD)}}. ACM,
  \bibinfo{pages}{1625--1631}.
\newblock


\bibitem[Shan et~al\mbox{.}(2016)]%
        {shan2016deep}
\bibfield{author}{\bibinfo{person}{Ying Shan}, \bibinfo{person}{T~Ryan Hoens},
  \bibinfo{person}{Jian Jiao}, \bibinfo{person}{Haijing Wang},
  \bibinfo{person}{Dong Yu}, {and} \bibinfo{person}{JC Mao}.}
  \bibinfo{year}{2016}\natexlab{}.
\newblock \showarticletitle{Deep crossing: Web-scale modeling without manually
  crafted combinatorial features}. In \bibinfo{booktitle}{\emph{Proceedings of
  the 22nd ACM SIGKDD International Conference on Knowledge Discovery and Data
  Mining (KDD)}}. ACM, \bibinfo{pages}{255--262}.
\newblock


\bibitem[Song et~al\mbox{.}(2019)]%
        {song2019autoint}
\bibfield{author}{\bibinfo{person}{Weiping Song}, \bibinfo{person}{Chence Shi},
  \bibinfo{person}{Zhiping Xiao}, \bibinfo{person}{Zhijian Duan},
  \bibinfo{person}{Yewen Xu}, \bibinfo{person}{Ming Zhang}, {and}
  \bibinfo{person}{Jian Tang}.} \bibinfo{year}{2019}\natexlab{}.
\newblock \showarticletitle{Autoint: Automatic feature interaction learning via
  self-attentive neural networks}. In \bibinfo{booktitle}{\emph{Proceedings of
  the 28th ACM International Conference on Information and Knowledge Management
  (CIKM)}}. ACM, \bibinfo{pages}{1161--1170}.
\newblock


\bibitem[Tian et~al\mbox{.}(2023)]%
        {tian2023eulernet}
\bibfield{author}{\bibinfo{person}{Zhen Tian}, \bibinfo{person}{Ting Bai},
  \bibinfo{person}{Wayne~Xin Zhao}, \bibinfo{person}{Ji-Rong Wen}, {and}
  \bibinfo{person}{Zhao Cao}.} \bibinfo{year}{2023}\natexlab{}.
\newblock \showarticletitle{EulerNet: Adaptive Feature Interaction Learning via
  Euler's Formula for CTR Prediction}. In \bibinfo{booktitle}{\emph{Proceedings
  of the 46th International ACM SIGIR Conference on Research and Development in
  Information Retrieval (SIGIR)}}. \bibinfo{pages}{1376--1385}.
\newblock


\bibitem[Wang et~al\mbox{.}(2024a)]%
        {wang2024flip}
\bibfield{author}{\bibinfo{person}{Hangyu Wang}, \bibinfo{person}{Jianghao
  Lin}, \bibinfo{person}{Xiangyang Li}, \bibinfo{person}{Bo Chen},
  \bibinfo{person}{Chenxu Zhu}, \bibinfo{person}{Ruiming Tang},
  \bibinfo{person}{Weinan Zhang}, {and} \bibinfo{person}{Yong Yu}.}
  \bibinfo{year}{2024}\natexlab{a}.
\newblock \showarticletitle{FLIP: Fine-grained Alignment between ID-based
  Models and Pretrained Language Models for CTR Prediction}. In
  \bibinfo{booktitle}{\emph{Proceedings of the 18th ACM Conference on
  Recommender Systems (RecSys)}}. \bibinfo{pages}{94--104}.
\newblock


\bibitem[Wang et~al\mbox{.}(2017)]%
        {wang2017deep}
\bibfield{author}{\bibinfo{person}{Ruoxi Wang}, \bibinfo{person}{Bin Fu},
  \bibinfo{person}{Gang Fu}, {and} \bibinfo{person}{Mingliang Wang}.}
  \bibinfo{year}{2017}\natexlab{}.
\newblock \showarticletitle{Deep \& cross network for ad click predictions}. In
  \bibinfo{booktitle}{\emph{Proceedings of the International Workshop on Data
  Mining for Online Advertising (ADKDD)}}. ACM, \bibinfo{pages}{12}.
\newblock


\bibitem[Wang et~al\mbox{.}(2021)]%
        {wang2021dcn}
\bibfield{author}{\bibinfo{person}{Ruoxi Wang}, \bibinfo{person}{Rakesh
  Shivanna}, \bibinfo{person}{Derek Cheng}, \bibinfo{person}{Sagar Jain},
  \bibinfo{person}{Dong Lin}, \bibinfo{person}{Lichan Hong}, {and}
  \bibinfo{person}{Ed Chi}.} \bibinfo{year}{2021}\natexlab{}.
\newblock \showarticletitle{Dcn v2: Improved deep \& cross network and
  practical lessons for web-scale learning to rank systems}. In
  \bibinfo{booktitle}{\emph{Proceedings of the Web Conference (WWW) 2021}}.
  \bibinfo{pages}{1785--1797}.
\newblock


\bibitem[Wang et~al\mbox{.}(2024b)]%
        {wang2024adsnet}
\bibfield{author}{\bibinfo{person}{Ruize Wang}, \bibinfo{person}{Hui Xu},
  \bibinfo{person}{Ying Cheng}, \bibinfo{person}{Qi He}, \bibinfo{person}{Xing
  Zhou}, \bibinfo{person}{Rui Feng}, \bibinfo{person}{Wei Xu},
  \bibinfo{person}{Lei Huang}, {and} \bibinfo{person}{Jie Jiang}.}
  \bibinfo{year}{2024}\natexlab{b}.
\newblock \showarticletitle{ADSNet: Cross-Domain LTV Prediction with an
  Adaptive Siamese Network in Advertising}. In
  \bibinfo{booktitle}{\emph{Proceedings of the 30th ACM SIGKDD Conference on
  Knowledge Discovery and Data Mining (KDD)}}. \bibinfo{pages}{5872--5881}.
\newblock


\bibitem[Yuan et~al\mbox{.}(2023)]%
        {yuan2023go}
\bibfield{author}{\bibinfo{person}{Zheng Yuan}, \bibinfo{person}{Fajie Yuan},
  \bibinfo{person}{Yu Song}, \bibinfo{person}{Youhua Li},
  \bibinfo{person}{Junchen Fu}, \bibinfo{person}{Fei Yang},
  \bibinfo{person}{Yunzhu Pan}, {and} \bibinfo{person}{Yongxin Ni}.}
  \bibinfo{year}{2023}\natexlab{}.
\newblock \showarticletitle{Where to go next for recommender systems? id-vs.
  modality-based recommender models revisited}. In
  \bibinfo{booktitle}{\emph{Proceedings of the 46th International ACM SIGIR
  Conference on Research and Development in Information Retrieval (SIGIR)}}.
  \bibinfo{pages}{2639--2649}.
\newblock


\bibitem[Zhang et~al\mbox{.}(2016)]%
        {zhang2016deep}
\bibfield{author}{\bibinfo{person}{Weinan Zhang}, \bibinfo{person}{Tianming
  Du}, {and} \bibinfo{person}{Jun Wang}.} \bibinfo{year}{2016}\natexlab{}.
\newblock \showarticletitle{Deep learning over multi-field categorical data}.
  In \bibinfo{booktitle}{\emph{European Conference on Information Retrieval
  (ECIR)}}. Springer, \bibinfo{pages}{45--57}.
\newblock


\bibitem[Zhang et~al\mbox{.}(2023)]%
        {zhang2023reformulating}
\bibfield{author}{\bibinfo{person}{Yang Zhang}, \bibinfo{person}{Tianhao Shi},
  \bibinfo{person}{Fuli Feng}, \bibinfo{person}{Wenjie Wang},
  \bibinfo{person}{Dingxian Wang}, \bibinfo{person}{Xiangnan He}, {and}
  \bibinfo{person}{Yongdong Zhang}.} \bibinfo{year}{2023}\natexlab{}.
\newblock \showarticletitle{Reformulating CTR Prediction: Learning Invariant
  Feature Interactions for Recommendation}. In
  \bibinfo{booktitle}{\emph{Proceedings of the 46th International ACM SIGIR
  Conference on Research and Development in Information Retrieval (SIGIR)}}.
  \bibinfo{pages}{1386--1395}.
\newblock


\bibitem[Zheng et~al\mbox{.}(2024)]%
        {zheng2024adapting}
\bibfield{author}{\bibinfo{person}{Bowen Zheng}, \bibinfo{person}{Yupeng Hou},
  \bibinfo{person}{Hongyu Lu}, \bibinfo{person}{Yu Chen},
  \bibinfo{person}{Wayne~Xin Zhao}, \bibinfo{person}{Ming Chen}, {and}
  \bibinfo{person}{Ji-Rong Wen}.} \bibinfo{year}{2024}\natexlab{}.
\newblock \showarticletitle{Adapting large language models by integrating
  collaborative semantics for recommendation}. In
  \bibinfo{booktitle}{\emph{2024 IEEE 40th International Conference on Data
  Engineering (ICDE)}}. IEEE, \bibinfo{pages}{1435--1448}.
\newblock


\bibitem[Zhou et~al\mbox{.}(2019)]%
        {zhou2019deep}
\bibfield{author}{\bibinfo{person}{Guorui Zhou}, \bibinfo{person}{Na Mou},
  \bibinfo{person}{Ying Fan}, \bibinfo{person}{Qi Pi}, \bibinfo{person}{Weijie
  Bian}, \bibinfo{person}{Chang Zhou}, \bibinfo{person}{Xiaoqiang Zhu}, {and}
  \bibinfo{person}{Kun Gai}.} \bibinfo{year}{2019}\natexlab{}.
\newblock \showarticletitle{Deep interest evolution network for click-through
  rate prediction}. In \bibinfo{booktitle}{\emph{Proceedings of the AAAI
  Conference on Artificial Intelligence (AAAI)}}, Vol.~\bibinfo{volume}{33}.
  \bibinfo{pages}{5941--5948}.
\newblock


\bibitem[Zhou et~al\mbox{.}(2018)]%
        {zhou2018deep}
\bibfield{author}{\bibinfo{person}{Guorui Zhou}, \bibinfo{person}{Xiaoqiang
  Zhu}, \bibinfo{person}{Chenru Song}, \bibinfo{person}{Ying Fan},
  \bibinfo{person}{Han Zhu}, \bibinfo{person}{Xiao Ma},
  \bibinfo{person}{Yanghui Yan}, \bibinfo{person}{Junqi Jin},
  \bibinfo{person}{Han Li}, {and} \bibinfo{person}{Kun Gai}.}
  \bibinfo{year}{2018}\natexlab{}.
\newblock \showarticletitle{Deep interest network for click-through rate
  prediction}. In \bibinfo{booktitle}{\emph{Proceedings of the 24th ACM SIGKDD
  International Conference on Knowledge Discovery \& Data Mining (KDD)}}. ACM,
  \bibinfo{pages}{1059--1068}.
\newblock


\end{thebibliography}

\end{document}